\RequirePackage{fix-cm}
\documentclass[smallextended]{svjour3}       
\smartqed  
\usepackage{graphicx}
\journalname{ }
\begin{document}

\title{Peculiarities of $^{57}$Fe NMR Spectrum in Micro- and Nanocrystalline Europium Orthoferrites
}

\titlerunning{$^{57}$Fe NMR of Europium Orthoferrites}        

\author{Anastasia~Sklyarova$^{1}$ \and Vadim~I.~Popkov$^{2}$ \and Ivan~V.~Pleshakov$^{3}$ \and Vladimir~V.~Matveev$^{4}$ \and Helena~\v{S}t\v{e}p\'ankov\'a$^{1}$ \and Vojt\v{e}ch~Chlan$^{1}$
}


\institute{A.~Sklyarova \\
              \email{asklyaro@gmail.com} \\
              $^1$ Faculty of Mathematics and Physics, Charles University, 18000 Prague 8, Czech Republic,\\
              $^2$ Hydrogen Energy Laboratory, Ioffe Institute, 194021 Saint Petersburg, Russia,\\
              $^3$ Laboratory of Quantum Electronics , Ioffe Institute, 194021 Saint Petersburg, Russia,\\
              $^4$ Faculty of Physics, Saint Petersburg State University, 198504 Peterhof, Saint Petersburg, Russia}

\date{ }

\maketitle

\begin{abstract}
NMR spectra of $^{57}$Fe dispersed europium orthoferrite in powder samples with micro- and nanocrystalline particles were studied for the first time. The material was synthesized by glycine-nitrate combustion, which allowed to obtain the specimens with granular diameters of 60~nm (nano-EuFeO$_3$) and 1.5~$\mu$m (micro-EuFeO$_3$). It was found out that the spectra are more complex than could be expected for a compound with a single crystallographic position of Fe$^{3+}$ ions, and it was also identified that there is a noticeable difference in samples with different fineness. Assumptions about the possible physical nature of the observed effects are made.
\keywords{Europium orthoferrites \and Nanomaterials \and Nuclear Magnetic Resonance}
\end{abstract}

\section{Introduction}
\label{intro}
Rare-earth orthoferrites in whole and  EuFeO$_3$, in particular, are subjected to a relentless interest due to their promising properties for practical use. Being potential materials for the finding of multiferroic properties, these ferrites undergo attention concerning their synthesis, modifications and study of their properties. Although EuFeO$_3$ is known already quite a long time and this  material in the bulk state has been well studied, since the beginning of nanomaterials era the interest to this substance has been renewed due to the potential application of its nanocrystals in photocatalytic materials, memory devices, gas sensors, etc  \cite{bib1,bib2,bib3,bib4}. 

It would seem that the properties of EuFeO$_3$ are well known but there is no clear information about the magnetic moments ordering and how it changes in the different material state (bulk, nano, thin film). Previous research results show that physical and chemical properties of ferrites may depend on the preparation route used for obtaining these materials, as it was shown for YFeO$_3$, where the spin reorientation transition was found in the hydrothermally-prepared samples \cite{bib5}. Moreover, some deviations on temperature curves of the magnetic susceptibility have been found for LuFeO$_3$ and EuFeO$_3$, prepared by hydrothermal method, which may indicate the existence of spin reorientation in these samples too \cite{bib6}, although the investigation of similar substances, prepared by other methods, contradicts the presence of spontaneous spin reorientation transition in these type of materials \cite{bib7}. Concerning the magnetic structure, \textit{AB}O$_3$ material shows the variety of spin orderings starting from ordinary antiferromagnetic (AFM) structure through different types of AFM, which may coexist with a weak ferromagnetism (WFM), to spiral or cycloidal distribution of spins \cite{bib8,bib9,bib10,bib11}. Type of spin ordering depends on many factors including a synthesis route, which influences on the physical properties of obtained materials and, that is most interesting, on the magnetic structure \cite{bib12,bib13}. 

In this work, we would like to discuss some features and differences of NMR spectra measured from nano- and microcrystalline EuFeO$_3$ powders in comparison with known facts about the magnetic structure of europium and some related orthoferrites.

\section{Results and Discussion}
\label{sec:1}
The stoichiometric glycine-nitrate combustion (\textit{GNC}) was used to produce europium orthoferrite; a detailed description of the synthesis procedure is given in \cite{bib14}. The controllable synthesis of nano- and microstructured powders of EuFeO$_3$ was carried out by the subsequent stabilizing (500$^\circ$C) or sintering (1000$^\circ$C) heat treatment of \textit{GNC} products, correspondently. The powders were then characterized by PXRD (\textit{Rigaku SmartLab 3}), EDX and SEM methods (\textit{Vega Tescan SBH} equipped with EDS by \textit{Oxford Instruments}).

\begin{figure}[htb]%
\includegraphics*[width=\linewidth]{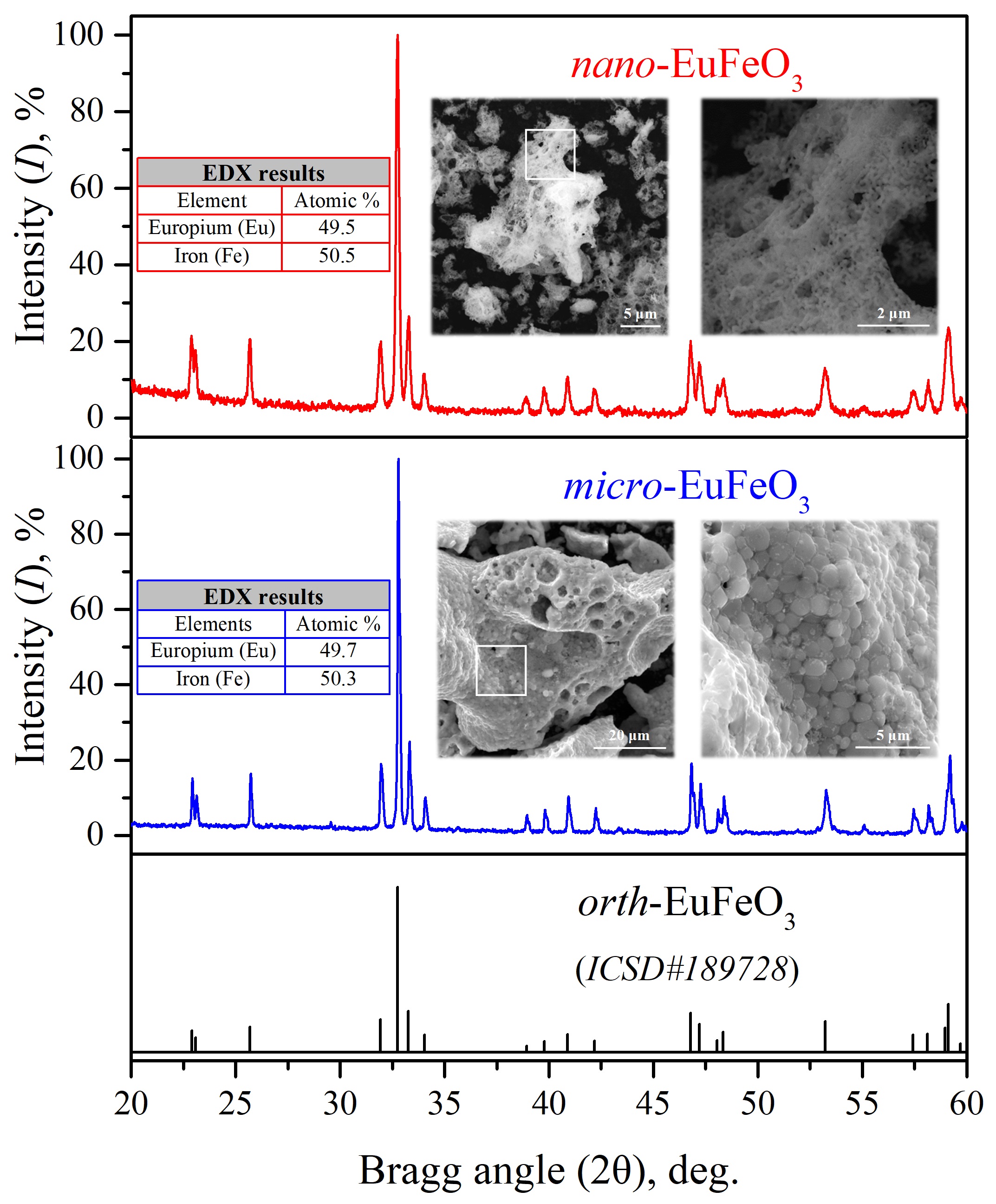}
\caption{PXRD patterns of the nano-EuFeO$_3$ and micro-EuFeO$_3$ powders. Table and image insets correspond to EDX and SEM results, correspondently}
\label{PXRD}
\end{figure}

The PXRD patterns of the nano-EuFeO$_3$ and micro-EuFeO$_3$ powders are represented in \textbf{Figure \ref{PXRD}}. It is shown that the phase composition of both samples fully corresponds to the phase-pure europium orthoferrite (ICSD card \# 189728). The XRD pattern of the nano-EuFeO$_3$ powders exhibits broadened peaks in the whole Bragg-angle range, indicating that europium orthoferrite is nanostructured. The sharp and clear diffraction peaks of the micro-EuFeO$_3$ powder are also indexed well to an orthorhombic europium orthoferrite, confirming the formation of highly crystalline micropowders. In this case, the sharper and stronger diffraction peaks demonstrate the enhanced crystallinity and particle size of the micro-EuFeO$_3$ powder in relation to nano-EuFeO$_3$.
The elemental analysis results of the synthesized samples of nano-EuFeO$_3$ and micro-EuFeO$_3$ show that the atomic fractions of europium (Eu) and iron (Fe) are equal to 49.5\%~:~50.5\%~at. and 49.7\%~:~50.3\%~at. (see table insets in Figure \ref{PXRD}), respectively, and indicate that the composition of the substances corresponds to the nominal composition of europium orthoferrite (50\%~:~50\%~at.) within the error of the method ($\pm$~1\%~at.). Then scanning electron microscopy was used to evaluate the morphological aspects of the obtained europium orthoferrite powders (see image insets in  Figure \ref{PXRD}). According to the presented results, nano-EuFeO$_3$ powder is characterized by an isometric morphology of particles with a size in the range of 40-80~nm (about 60~nm or 0.06~$\mu$m in average), which are agglomerated into foam-like micron structures that are characteristic of REE orthoferrites in case of the synthesis by the solution combustion method \cite{bib15,bib16}. A similar situation is observed for the micro-EuFeO$_3$ powder with the only difference being that the characteristic sizes of the isometric particles of europium orthoferrite are about 1-2~$\mu$m (about 1.5~$\mu$m or 1500~nm in average), but they are also agglomerated into larger structures that retain the foam-like motif of the precursor observed previously \cite{bib17}. So, both SEM and EDX results are in a good agreement with the PXRD data, confirming EuFeO$_3$ particles are monocrystalline in both nano- and micropowder.

Thus, the main difference between the nano-EuFeO$_3$ and micro-EuFeO$_3$ samples is in the average particle size of the europium orthoferrite particles equal to of 0.06~$\mu$m (60~nm) and 1.5~$\mu$m (1500~nm) with the same chemical and phase composition, crystal structure, morphology and particle size distribution. Therefore, the observed features and differences of the $^{57}$Fe NMR spectra of these samples can be associated only with the influence of the size factor.

Earlier works devoted to magnetic properties study of  \textit{AB}O$_3$ (\textit{A} - rare-earth element (REE), \textit{B} = Mn, Fe)  show that, with temperature changing, the Fe-sublattice in this structure undergoes one or several magnetic transitions, but, concerning the rare-earth sublattice, experimental data confirming the existence of magnetic transition below 10~K and  refuting one are contradicted each other \cite{bib18,bib19,bib20,bib21,bib22} . The dependence of ferrite properties, including phase transitions, on preparation way is evident and the result is unpredictable, but, at the same time, magnetic structure is an important parameter in multiferroics, where magnetoelectric effect is associated with phase transitions and emerging magnetic order. 

For obtaining information about the magnetic structure of micro- and nano-EuFeO$_3$ samples, $^{57}$Fe NMR spectra were acquired  at different temperatures starting from 4.2~K (\textit{Bruker Avance II} console modified for magnetic materials study).
\begin{figure}[htb]%
\includegraphics*[width=\linewidth]{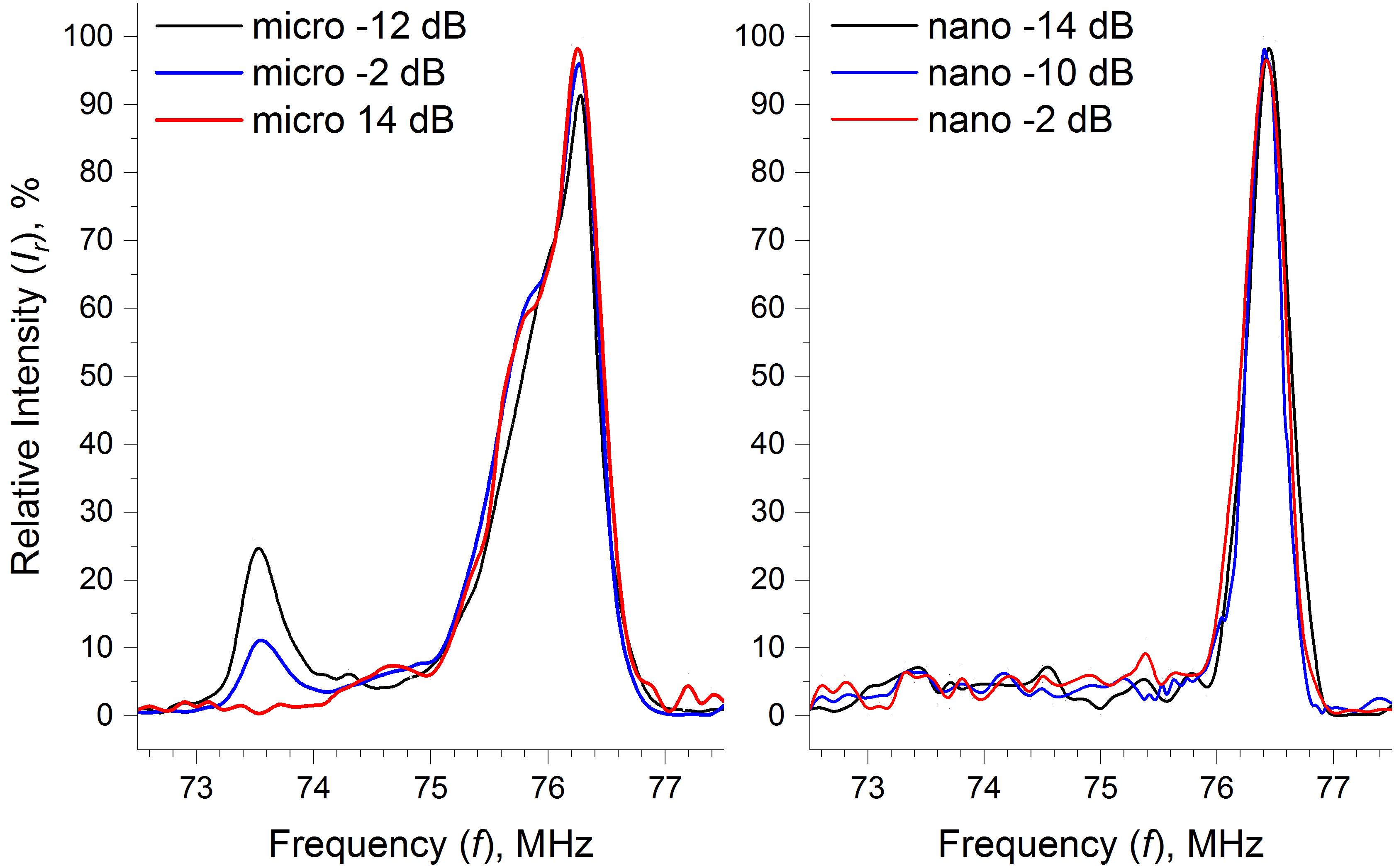}
\caption{$^{57}$Fe NMR spectra of the micro-EuFeO$_3$ and nano-EuFeO$_3$ powders acquired at 4.2~K and different RF power. All spectra are normalized to a maximal value of echo amplitude of the related sample.}
\label{NMR}
\end{figure}
As it is seen from \textbf{Figure \ref{NMR}},  NMR spectra of our samples have somewhat unusual shape: the spectrum of nanomaterial shows an almost single line, whereas the spectrum shape of micromaterial has a complex structure with more pronounced "shoulder". Observed spectral shapes are in contradiction with known works compared micro and nanomaterials, where surface effects lead to the significant complication of "nano" spectrum. But in our case the opposite picture was observed, which can be explained by the samples preparation technology as well as other possible effects described below. Also for nano-EuFeO$_3$ the main resonance line shows an $\sim$0.15~\% shift to higher frequency indicating the volume magnetization grow to the same value ($\sim$0.15~\%), which is typical for nanophase.  

\begin{figure}[htb]%
\includegraphics*[width=\linewidth]{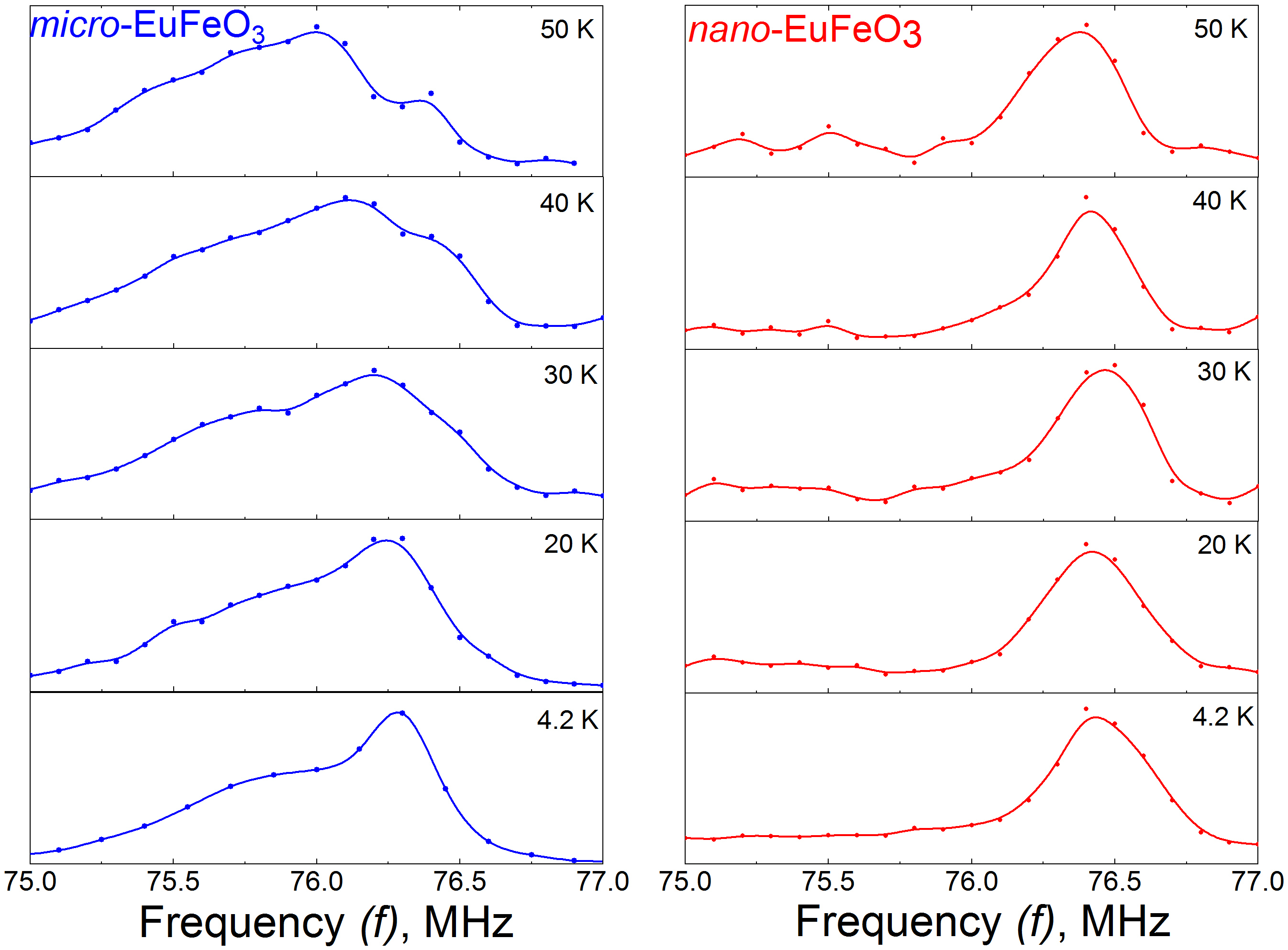}
\caption{$^{57}$Fe NMR spectra of micro-EuFeO$_3$ and nano-EuFeO$_3$ samples acquired at indicated temperatures.
}
\label{NMR_T}
\end{figure}

Increasing the temperature of NMR measurements gives the further complication of micro-EuFeO$_3$ spectrum, \textbf{Figure \ref{NMR_T}}, left panel. 
Between 20 and 30~K the main NMR peak starts to split and the almost separate line becomes visible by naked eye already at 50~K. Earlier, such temperature splitting of the main NMR line was observed in pure and substituted orthoferrites and was bonded with the spin reorientation transition in these materials \cite{bib23}. Thus, the same reason maybe adapts for the explanation of the line separation for micro-EuFeO$_3$.  

Origins of satellites in the micro-sample have been analyzed using NMR signal dependence on the power of radio-frequency (RF) pulses. For both samples, the main peak and "shoulder" show the similar RF dependence in the wide power-value region and can be attributed to europium orthoferrite, but the satellite peak at $\sim$73.5~MHz shows another dependence on RF power (see Figure \ref{NMR}, left panel). Based on the known equations for the behavior of echo signal versus pulse amplitude and taking into account amplification coefficient, low-frequency satellite has an one and a half times lower value of anisotropy field than the main peak and may be ascribed to another phase. Perhaps, this impurity phase is the maghemite or another oxide, the appearing of which is not excluded by the synthesis technology. This phase should be extremely fine to become invisible on the XRD pattern. 
 
In theory, upon the condition of homogeneous distribution of Fe atoms in orthorhombic structure, when Fe atoms occupy only one crystallographic position, $^{57}$Fe NMR spectrum of EuFeO$_3$ at 4.2~K should be presented by one narrow line. And, although the NMR resonance peak of nano-EuFeO$_3$ can be considered as a single line, it is possible to notice the presence of a weak "shoulder" as in micro-sample (Figure \ref{NMR}, right panel), that excludes the simple explanation of "shoulder" by the theory of domain walls. This conclusion is also confirmed by the conservation of "shoulder" in large enough external magnetic fields up to 2.6~T. It means, that other mechanisms of the appearing of complex spectrum structure should be considered, but all of them are bonded with a non-homogeneous distribution of magnetic moments in the sample volume.

One of the famous features of \textit{AB}O$_3$ compounds, showed multiferroic properties, is a spiral or cycloid spin ordering \cite{bib10}. But in ferrites this ordering was found only in BiFeO$_3$ multiferroics, where the complex NMR spectrum was explained by spin modulated magnetic structure: cycloid spin distribution \cite{bib24,bib25,bib26}. Because of this spin distribution, low temperature NMR spectrum of bulk BiFeO$_3$ consists of wide ($\sim$1~MHz) line with two maxima \cite{bib26}. As it was shown in a number of works, this cycloida is supressed by application of a strong external magnetic field, by producing strained thing films of this material, by substituting of Bi by rare-earth elements, and other methods \cite{bib27,bib28,bib29,bib30,bib31}. Earlier, $^{57}$Fe MNR spectra have been obtained for La doped Bi$_{1-x}$La$_x$FeO$_3$ and it was shown that at $x=0.3$ NMR line has only one broad peak; at $x\geq0.9$ this peak becomes narrow and shifts to high frequencies \cite{bib32}. This effect was explained by the rhombohedral-to-orthorhombic phase transition accompanied by the spin cycloid destruction. At substitution of Bi by Eu in Bi$_{1-x}$Eu$_x$FeO$_3$ ($x\geq0.3$) the same phase transition was observed \cite {bib33,bib34} and it would be reasonable to expect the same behavior of NMR spectrum in the whole substitution range up to $x=1$. However, the spectrum shape of our micro-EuFeO$_3$ sample looks rather similar to the spin distribution case, moreover, collapse of the NMR spectrum in one narrow line with the particle size reducing to nanometer range also looks rather similar to behavior of nano-BiFeO$_3$, where cycloid is not present due to the fact that the particle size is smaller than the period of $\sim$62~nm \cite{bib35}. Nonetheless, in rare-earth element $\it{A}$MnO$_3$ multiferroics the coexistence of orthorhombic phase and spin modulated structure is not forbidden \cite{bib10,bib36}, and even for Y-doped EuMnO$_3$ the spiral spin distribution was observed due to the formation of distortions in the orthorhombic $\it{Pbnm}$ structure \cite{bib9,bib37}. But in ferrites, only other phases were found to be suitable for stabilization of the spin-modulated structure and similarity of our NMR spectra and spectra of BiFeO$_3$ may be formal. 

Another hypothesis to explain our results is the presence of ferroelectric domain borders, which may also be responsible for the non-homogeneous distribution of magnetic moments. In this case, the observed complex spectrum shape of micro-material is explained by the big crystallites size, which may exceed the domain size. The existence of multiferroic properties in rare-earth \textit{A}FeO$_3$ orthoferrites  was confirmed by a number of works \cite{bib38}. In these works, the appearance of room temperature ferroelectricity in centrosymmetric ferrites was explained by two nonequivalent canted antiferromagnetic subsystems and was called "spin-canting-driven ferroelectricity" \cite{bib39}.  EuFeO$_3$ has an AFM ordering with a weak ferromagnetism and quite satisfies the conditions of the ferroelectricity emergence.

\section{Conclusions}
\label{sec:3}
The described NMR study of micro- and nano-powders of EuFeO$_3$ shows the appearing of new properties in europium orthoferrites produced by \textit{GNC} method. Obtained results point to possible SR transition in the absence of an external field and the appearing of ferroelectric properties in the samples. We believe that further thorough investigation of obtained effects may give important information about the mechanism of multiferroic properties formation in rare-earth orthoferrites.

\begin{acknowledgements}
The research was supported by OP RDE project \\
No. CZ.02.2.69$\slash$0.0$\slash$0.0$\slash$16$\_$027$\slash$0008495,\\
International Mobility of Researchers at Charles University.
\end{acknowledgements}

\end{document}